# Second Chern crystals in four-dimensional synthetic translation space with inherently nontrivial topology

Xiao-Dong Chen, Fu-Long Shi, Jian-Wei Liu, Ke Shen, Xin-Tao He, Wen-Jie Chen[*], Jian-Wen Dong[*]

*School of Physics & State Key Laboratory of Optoelectronic Materials and Technologies, Sun Yat-sen University, Guangzhou 510275, China.*

[*]Corresponding author: chenwenj5@mail.sysu.edu.cn, dongjwen@mail.sysu.edu.cn

## ABSTRACT

**Topological states, first known as quantum Hall effect or Chern insulating crystal, have been generalized to many classical wave systems where potential applications such as robust waveguiding, quantum computing and high-performance lasers are expected. However, a crystal can be either topologically trivial or nontrivial, depending on its detailed configuration, and one needs to carefully design the structure and calculate its topological invariant before the actual applications. Here, we theoretically study and experimentally demonstrate the second Chern crystal in a four-dimensional space by introducing two extra synthetic translation dimensions. Due to the inherently nontrivial topology of the synthetic translation space, this abstract four-dimensional crystal is guaranteed to be topologically nontrivial regardless of the detailed configuration. The dimensional hierarchy of gapless boundary modes can be deduced by dimension reduction. Remarkably, one-dimensional gapless dislocation modes are observed and their robustness is confirmed in our experiments. This ubiquitous phenomenon in synthetic translation space provides perspectives on the findings of topologically nontrivial crystals and inspires the designs of classical wave devices.**

Topological states of matter and their extensions to classical wave systems have revolutionized our understanding of the boundary characteristic of a crystal [1-5]. The boundary properties have a strong correspondence with the band topology of bulk crystal states. In this context, the bulk topology can be attributed as an eigenvalue problem in the Bloch momentum space. And the so-called band topology is, essentially, the connection between the eigen wavefunctions within the Bloch momentum space and characterized by global topological invariants. One prominent example is the first Chern number defined in two-dimensional (2D) momentum space, which counts the phase winding of wavefunction along Brillouin zone boundary [6, 7]. The band insulators with nonzero Chern number are referred to as Chern insulators which support chiral boundary states and quantized Hall conductance. Generally speaking, the Chern number and band topology would depend on the system parameters and can be any integer (trivial or nontrivial) in principle.

In addition, the parameter space of the eigenvalue problem, where the band topology is concerned, is not restricted to the momentum space. By introducing the concept of synthetic dimensions, one can discuss the dependency of wavefunction on any system parameter (applied external fields, constitutive parameters, geometric sizes, etc.) and investigate the band topology in synthetic space [8-11]. One obvious advantage to do so is that in this general synthetic space, one can study the higher dimensional topological states and go beyond the physical dimensionality of real space. A prototypical example is the 4D Chern insulators [12, 13], i.e. the 4D generalization of 2D quantum Hall effect, by constructing a 4D lattice in synthetic space. The 4D band topology is accordingly characterized by the second Chern number and would be manifested in the gapless boundary modes on lower dimensions and the quantized Hall response. For the high design flexibility and the easier fabrication of sample, classical wave topological materials [14-18] are good platforms to study the higher dimensional topological physics in synthetic space and further explore their potential application [19-25].

Here, we investigate and realize second Chern crystals in a 4D synthetic translation space which is spanned by a 2D Bloch momentum space and a 2D translation parameter space. Different from the aforementioned synthetic space, the band topology in this translation space is guaranteed to be nontrivial with a second Chern number of +1 due to the unique connection between wavefunctions. By exploiting the inherent nontrivial topology, we illustrate the gapless 2D surface modes which are well expected with dimension reduction from the 4D synthetic translation space. In addition, we experimentally observe the existence and robustness of 1D vortex line modes which manifest as 0D dislocation modes in real space. This phenomenon ensured by topological band theory will undoubtedly renovate our design philosophy of photonic devices and may pave the way to novel classical wave devices based on the synthetic translation space.

We consider a typical 2D PC composed of a square lattice of dielectric rods, and all rods are translated away from the origin by ($\Delta x$, $\Delta y$) [Fig. 1(a)]. Combining this 2D translation space and the 2D Bloch momentum space, we can construct a 4D synthetic space [Fig. 1(c)]. Figure 1(b) illustrates bulk bands of the transverse magnetic modes of PC with ($\Delta x$, $\Delta y$) = (0, 0) and shows a complete band gap spanning from 7.24 to 9.63 GHz. As PCs with different ($\Delta x$, $\Delta y$) have the same bulk bands, a 4D band gap naturally exists in the ($k_x$, $\Delta x$, $k_y$, $\Delta y$) space, and the second Chern number of the lowest bulk band is well-defined:

$$C^{(2)} = \frac{1}{32\pi^2} \int \varepsilon_{lmno} B_{lm}(k_x, \Delta x, k_y, \Delta y) B_{no}(k_x, \Delta x, k_y, \Delta y) dk_x d\Delta x dk_y d\Delta y, \qquad (1)$$

where $\varepsilon_{lmno}$ is the Levi-Civita symbol, Berry curvature $B_{lm}(\mathbf{k}) = \partial_l A_m(\mathbf{k}) - \partial_m A_l(\mathbf{k})$ is written in terms of the Berry connection $A_l(\mathbf{k}) = \langle u(\mathbf{k}) | \partial_l | u(\mathbf{k}) \rangle$. Here, $u(\mathbf{k})$ is the periodic wavefunction and each index $l$, $m$, $n$, $o$ takes values of $k_x$, $\Delta x$, $k_y$, or $\Delta y$. Considering the symmetries of the Berry curvature, i.e., $B_{lm} = -B_{ml}$, the second Chern number can be rewritten as $C^{(2)} = C^{(1)}_{k_x,\Delta x} C^{(1)}_{k_y,\Delta y} - C^{(1)}_{k_x,\Delta y} C^{(1)}_{k_y,\Delta x} - C^{(1)}_{k_x,k_y} C^{(1)}_{\Delta x,\Delta y}$ [26]. Therefore, we first calculate the first Chern numbers in the six 2D subspaces before getting the second Chern number. Figure 1(d) plots the Berry curvature distribution in the ($k_x$, $\Delta x$) and ($k_y$, $\Delta y$) subspaces, whose integral over the Brillouin zone ($C^{(1)}_{k_x,\Delta x}$ and $C^{(1)}_{k_y,\Delta y}$) are both -1. The $\Delta x$-independent ($\Delta y$-independent) Berry curvatures on the left (right) panel are distinctly different from those of 2D magnetic PCs [refs: MPC PRL08 Nature09], and find more comparison in Supplement A. Note that the first Chern numbers in these two subspaces are inherently nontrivial and must be -1, because the wavefunctions have fixed relations due to the translation. Meanwhile, $C^{(1)}_{k_x,\Delta y}$, $C^{(1)}_{k_y,\Delta x}$ and $C^{(1)}_{\Delta x,\Delta y}$ must be 0 (inherently trivial), which do not depend on the specific form of the 2D wavefunctions. In addition, $C^{(1)}_{k_x,k_y}$ is zero under the time reversal symmetry. Hence the second Chern number must be [see detailed derivation in Supplement B]:

$$C^{(2)} \equiv (-1) \times (-1) - (0) \times (0) - (0) \times (0) \equiv 1. \qquad (2)$$

In this sense, any 2D crystal with a complete band gap is a second Chern crystal in this unique 4D ($k_x$, $\Delta x$, $k_y$, $\Delta y$) space.

The inherently nontrivial band topology of this 4D space would manifest itself as the gapless boundary modes when the 4D bulk crystal is truncated to lower dimensional boundaries. As a typical example, 1D gapless dislocation modes can be obtained by truncating the 4D crystal in three dimensions [Fig. 1(e)], which will be clarified in detail below. Topological boundary modes also exist when the 4D crystal is truncated to a 3D hypersurface (leading to a 1D edge modes in real space which is presented in Supplement C), or to a 2D surface (leading to a 0D localized corner modes in real space). Here we truncate the crystal in both the *x* and *y* directions and form a square corner bounded by the perfect electric conductor (PEC), as illustrative in Fig. 2(a). Figure 2(b) calculates the dispersion of 2D surface modes (shaded in colors) in the synthetic ($\Delta x$, $\Delta y$) space. Note that the 3D hypersurface modes for the left and the bottom PEC boundaries projected onto the ($\Delta x$, $\Delta y$) plane are shaded in red and blue, respectively. Outside these two projected hypersurface bands are a set of gapless surface modes localized at the corner, which are protected by two subspace Chern numbers $C^{(1)}_{k_x,\Delta x} = C^{(1)}_{k_y,\Delta y} = -1$. To be concrete, we take the cut plane at $\Delta x = -0.3a$ for example [Fig. 2(c)]. The 0D corner mode (2D surface mode) can be viewed as the end mode of the 1D edge mode (3D hypersurface mode). And since the first Chern number of hypersurface modes at the left boundary is -1, gapless corner modes (colored curve) naturally exist above the projected hypersurface band (red). This argument also applies for a $\Delta y$ cut plane and therefore the 2D dispersion of the corner state is an arched sheet [colored in Fig. 2(b)] connecting the two projected hypersurface bands [red and blue in Fig. 2(b)]. The field pattern $|E_z|$ of one representative corner mode at ($\Delta x$, $\Delta y$) = (-0.3$a$, -0.2$a$) is shown in Fig. 2(d), showing the strong field confinement around the corner. Note that these gapless boundary modes always exist on the interface between the 4D crystals and trivial insulators [Supplement D], and the PEC is used as a representative example of trivial insulators.

Another manifestation of the nontrivial topology of a second Chern crystal is the gapless chiral dislocation modes along a 1D vortex line [27, 28]. To see this, we first fabricate a microwave sample where the translation of each rod away from its lattice point is (0.5$a$, $\theta a/2\pi$) [Fig. 3(a)]. Here $\Delta x$ is a constant while $\Delta y$ varies as the azimuthal angle $\theta$, leading to a dislocation structure near the origin with a Burgers vector of $\mathbf{B} = (0, a)$. Numerical simulation in Fig. 3(c) shows that such a dislocation supports an in-gap mode localized around the origin at $f = 9.2$ GHz. To demonstrate this, we carry out the near-field scanning measurement [Fig. 3(b)]. A source antenna is inserted through the metallic substrate to excite the eigenmodes and a probe antenna is mounted on a translational stage to measure the spatial field distributions. The measured field pattern $|E_z|$ of dislocation mode has a strong field localization around the origin [Fig. 3(d)], which is in good agreement with the simulated pattern. Such dislocation modes are gapless and chiral along the synthetic $\Delta x$ dimension when a 1D vortex line is considered [Fig. 3(e)]. Along the vortex line, five representative dislocation lattices (with $\Delta x$ =-0.5$a$, -0.25$a$, 0, 0.25$a$, 0.5$a$) are plotted as examples. For each lattice with a fixed $\Delta x$, a vortex point exists at the origin, around which are rods with different $\Delta y$. To preserve the continuity of lattices with $\Delta x$ = 0.5$a$ and $\Delta x$ = -0.5$a$, the diameter of the first rod to the left of the origin (outlined in red) is set as $d_0$ = -($\Delta x/a$ - 0.5)$d$. Figure 3(f) plots the eigen frequency spectra of in-gap dislocation modes as a function of $\Delta x$ [highlighted in cyan]. It exhibits a gapless dispersion traversing the whole band gap. Since $\Delta x$ plays the role of pseudo momentum along the vortex line, the eigen frequency of dislocation mode increases as $\Delta x$, which mimics a positive group velocity and is consistent with the second Chern number of +1. By using the pump-probe transmission method [refs], measured frequency dispersion of the dislocation modes (bright color) agrees well with the calculated one [Fig. 3(g)]. Note that the number of and the dispersion direction of gapless dislocation modes are in correspondence with the Burgers vector and one can realize more dislocation bands by considering a dislocation lattice with a longer Burgers vector [see more examples in Supplement E].

It is noteworthy that the topologically-protected cavity modes or corner modes have been investigated recently in various systems [29-31]. These topological mid-gap modes are robust against disorder to some extent or they are protected by artificial symmetries. But in principle, the eigen frequencies of these modes would inevitably merge into the bulk band as long as the perturbation is large enough or breaks the artificial symmetry. Similarly, this is true for our system, if we only focus on one lattice with a fixed $\Delta x$. But if we examine the whole spectrum as a function of $\Delta x$, we will find that gapless dislocation modes always span the nontrivial 4D band gap, and the number of these states is consistent with the second Chern number. To see this, we consider the same dislocation lattice as that in Fig. 3(a) but with an additional dielectric rod with the diameter of 5.6 mm or a rhombus PEC block with the side length of 9 mm [Figs. 4(a) and 4(d)]. Centers of the additional rod and PEC block are both fixed at (5 mm, 0) with respect to the origin. In both cases with introduced defects, the eigen frequency of the dislocation mode for $\Delta x = 0.5a$ is shifted from 9.2 to 9.75/9.83 GHz, merging into the bulk. But from the overall view of the spectra in Figs. 4(b) and 4(e), one can find that there always be a gapless band with its frequency increasing as $\Delta x$. In both defected cases, although the spectra are altered, the gapless feature is preserved by the nontrivial topology of the 4D synthetic space. This is also confirmed by our experimental results [Figs. 4(c) and 4(f)].

In summary, we have realized the second Chern crystals in the 4D synthetic translation space which is spanned by a 2D Bloch momentum space and a 2D translation parameter space. The inherent nontrivial band topology within this translation space guarantees the existence of chiral boundary modes (e.g. 2D surface modes, and 1D vortex line modes), which can be comprehended via dimension reduction. In particular, 1D vortex line modes manifested as 0D dislocation modes in real space were experimentally observed, as well as their robustness against defect. The existence of these lower dimensional boundary modes stems from the nontrivial topology of the unique parameter space $(k_x, \Delta x, k_y, \Delta y)$, rather than the specific parameters of the crystals. Hence this phenomenon is universal for crystals with any lattice, unit cell geometry or material parameter, as long as a complete band gap exists. The example of the crystal with triangular lattice, transverse electric polarization, unit cell consisting of air rod in dielectric background is shown in Supplement F. The ubiquitous phenomenon in synthetic translation space provides perspectives on the findings of topologically nontrivial crystals and has the potential applications in designs of topological nanocavities.

**Acknowledgments**

We thank Prof. Jian-Hua Jiang for the helpful discussions. This work was supported by National Natural Science Foundation of China (Grant Nos. 62035016, 12074443, 11874435, 11904421), the Young Top-Notch Talent for Ten Thousand Talent Program (2020-2023), Guangdong Basic and Applied Basic Research Foundation (Grant No. 2019B151502036), Natural Science Foundation of Guangdong Province (Grant No. 2018B030308005), Guangzhou Science, Technology and Innovation Commission (Grant Nos. 201904010223, and 202102020693), Fundamental Research Funds for the Central Universities (Grant No. 20lgzd29, 20lgjc05, 2021qntd27).

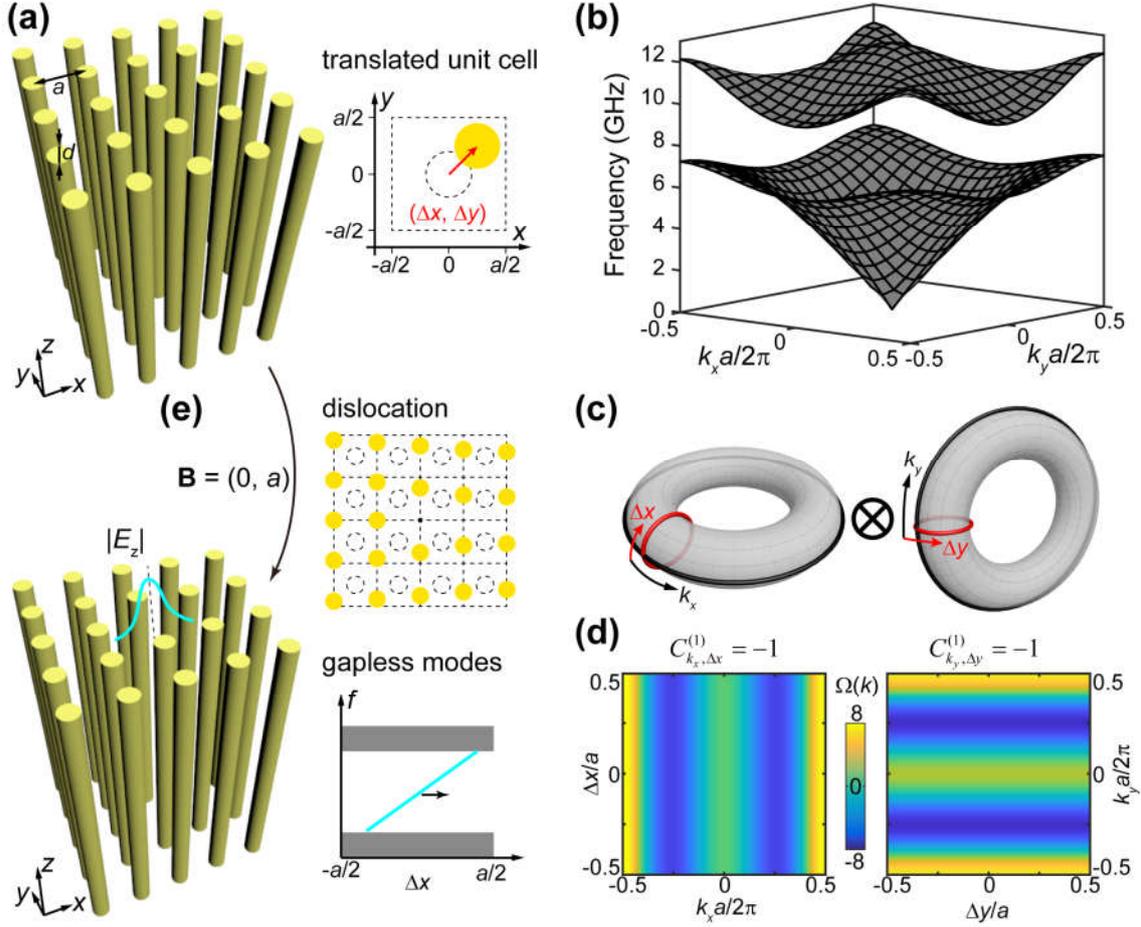

**FIG. 1. Second Chern crystal in 4D synthetic translation space and its gapless chiral dislocation modes.** (a) Schematic of a perfect lattice of dielectric rods (yellow) in the air. Inset: a translated unit cell where the rod is translated by ($\Delta x$, $\Delta y$) from the origin. Parameters: lattice constant $a$ = 14 mm, diameter and relative permittivity of rod $d$ = 5.6 mm and $\varepsilon$ = 8.0. (b) Bulk bands of the 2D PC with ($\Delta x$, $\Delta y$) = (0, 0) in the reciprocal space ($k_x$, $k_y$), along with a complete band gap spanning from 7.24 to 9.63 GHz. (c) Schematic of the 4D Brillouin zone spanned by ($k_x$, $\Delta x$, $k_y$, $\Delta y$). (d) Berry curvatures of the lowest bulk band in the ($k_x$, $\Delta x$) space with $k_y$ = 0 & $\Delta y$ = 0 (left) and in the ($k_y$, $\Delta y$) space with $k_x$ = 0.8$\pi$/$a$ & $\Delta x$ = -0.3$a$ (right), giving nonzero first Chern numbers in the corresponding 2D subspaces. (e) Schematic of the dislocation lattice characterized by Burgers vector of **B** = (0, $a$). A gapless dislocation mode is expected to traverse the bulk gap.

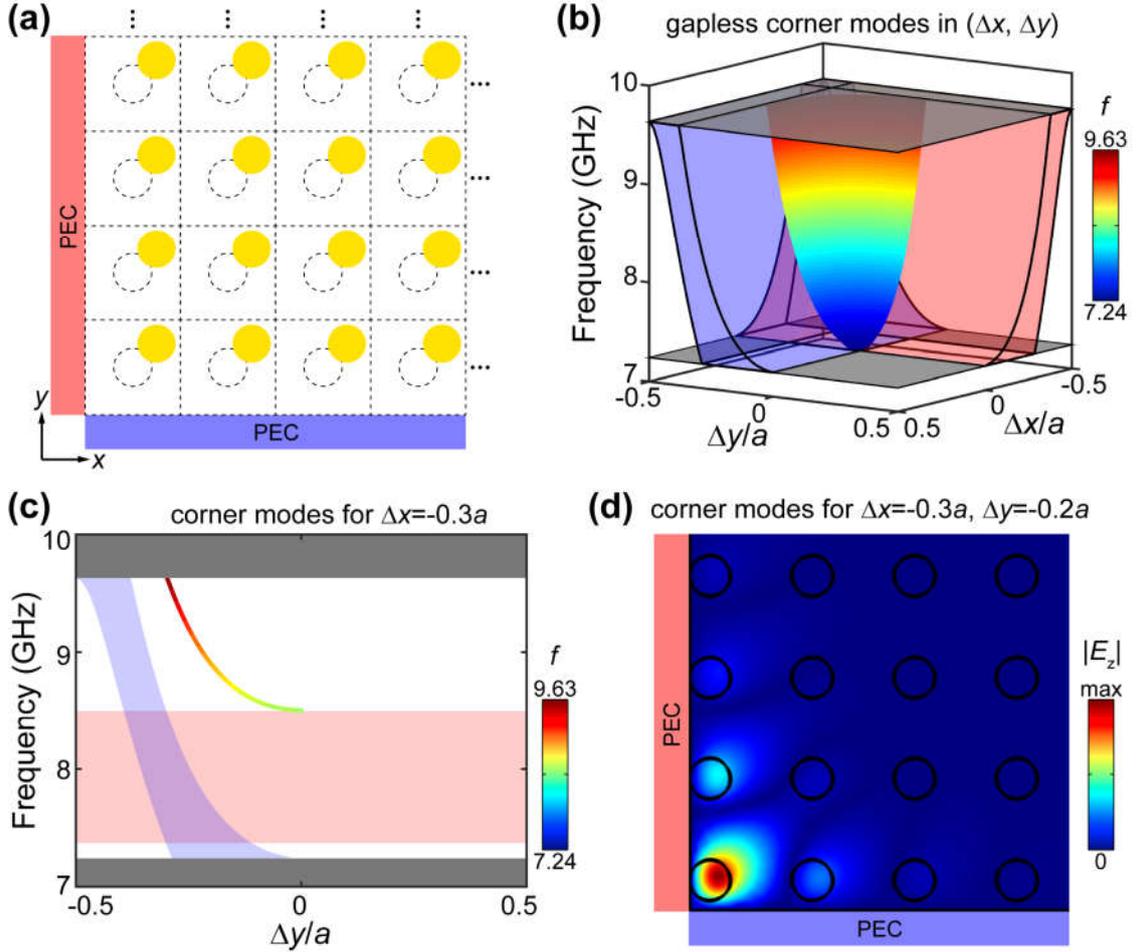

**FIG. 2. 2D topological surface modes manifested as 0D corner modes in real space.** (a) Schematic of the corner between the 2D PC with ($\Delta x$, $\Delta y$) and two PECs on the left and the bottom. (b) Gapless corner modes (shaded in colors) in the ($\Delta x$, $\Delta y$) space. The projected hypersurface bands of the two PEC boundaries are shaded in red and blue. Two transparent gray squares mark the lower and upper frequency edges of the band gap. (c) 1D gapless dispersion of corner modes of the second Chern crystal as a function of $\Delta y$, by fixing $\Delta x = -0.3a$. (d) Field pattern $|E_z|$ of the corner mode for the second Chern crystal with ($\Delta x$, $\Delta y$) = (-0.3$a$, -0.2$a$).

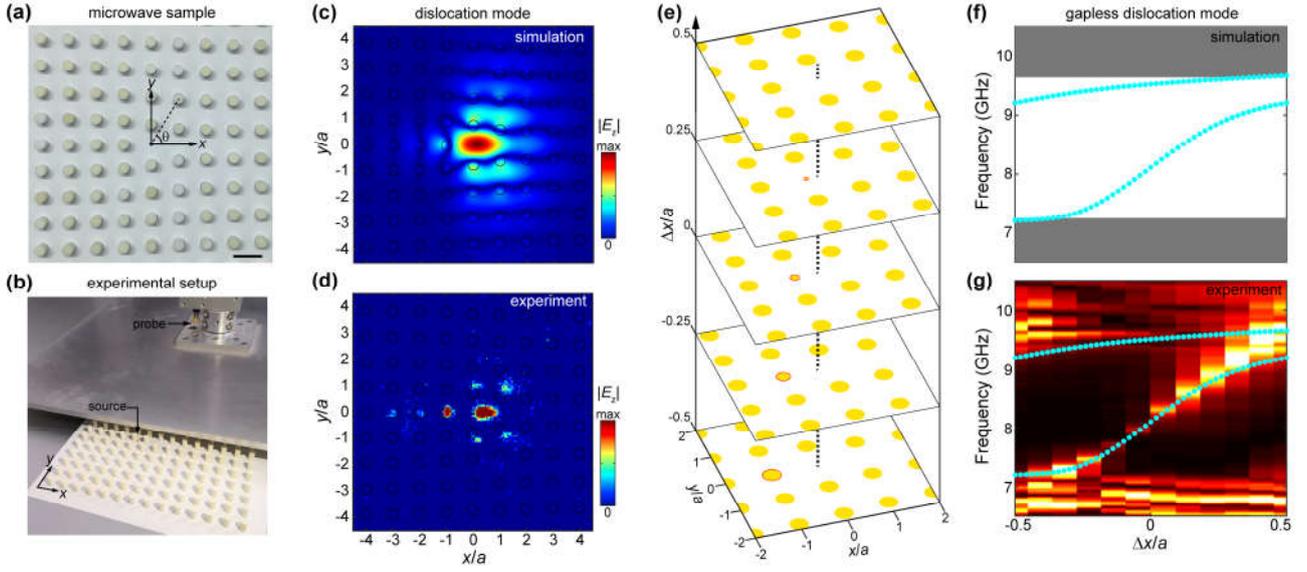

**FIG. 3. 1D vortex line modes manifested as 0D dislocation modes in real space.** (a) Microwave sample of the dislocation lattice. The translations of rods obeys $(0.5a, \theta a/2\pi)$, leading to a dislocation lattice with the Burgers vector of **B** = $(0, a)$. Scale bar: 14 mm. (b) Setup for the near-field scanning measurement. The top metallic plate is shifted to show the inside PC. (c-d) Simulated and measured field patterns $|E_z|$ of the dislocation mode. (e) 3D schematic of the 1D vortex line (dashed line) which is formed along the synthetic $\Delta x$ dimension. Five representative dislocation lattices (with $\Delta x$ =-0.5$a$, -0.25$a$, 0, 0.25$a$, 0.5$a$) are plotted as examples. To preserve the continuity of lattices with $\Delta x$ = 0.5$a$ and $\Delta x$ = -0.5$a$, the diameter of the first rod to the left of the origin (outlined in red) is set as $d_0 = -(\Delta x/a - 0.5)d$. (f-g) Simulated and measured eigenfrequency spectra in which gapless dislocation bands are observed.

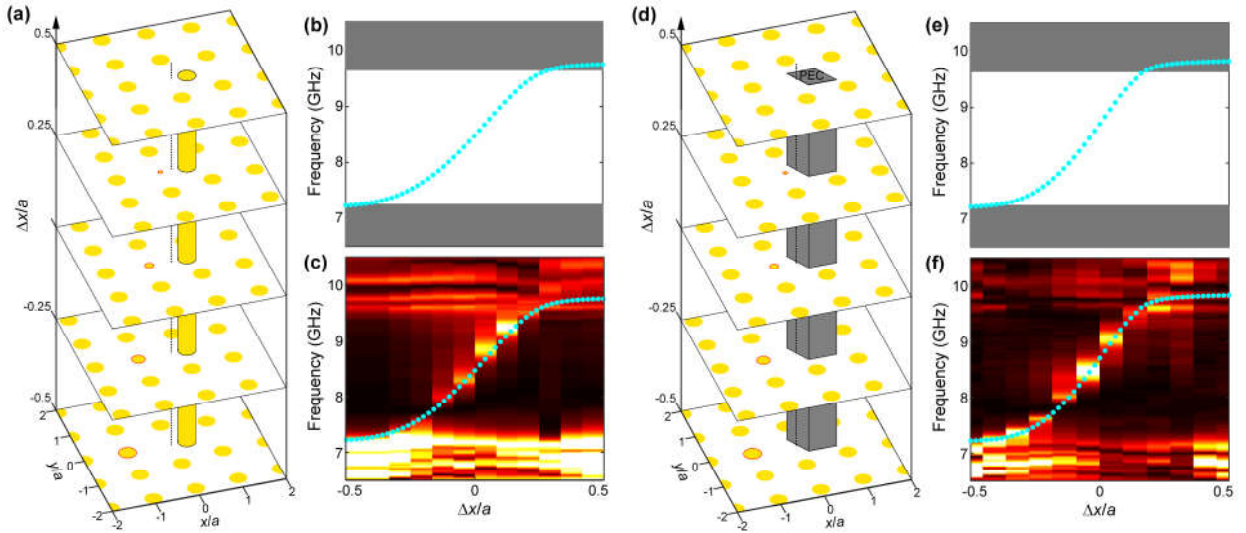

**FIG. 4. Robustness of gapless dislocation modes.** (a & d) 3D schematic of the 1D vortex line with (a) an additional dielectric rod and (d) a PEC defect on the right of the origin. (b-c, e-f) Simulated and measured frequency spectra for (b-c) the samples with an additional dielectric rod and (e-f) the samples with a PEC defect. In both cases, although the band dispersions are altered, the gapless chiral feature is preserved.

13